# High Resolution Microimaging with Pulsed Electrically-Detected Magnetic Resonance

Itai Katz,[*] Matthias Fehr,[#] Alexander Schnegg,[#] Klaus Lips,[#] and Aharon Blank[*1]

[*]Schulich Faculty of Chemistry
Technion – Israel Institute of Technology
Haifa, 32000
Israel

[#]Helmholtz-Zentrum Berlin für Materialien und Energie, Institut für Silizium-Photovoltaik,
Kekuléstr. 5, D-12489 Berlin
Germany

---

[1] Corresponding author contact details:   Aharon Blank, Schulich Faculty of Chemistry, Technion – Israel Institute of Technology, Haifa, 32000, Israel, phone: +972-4-829-3679, fax: +972-4-829-5948, e-mail: ab359@tx.technion.ac.il.

# Abstract

The investigation of paramagnetic species (such as point defects, dopants, and impurities) in solid-state electronic devices is significant because of their effect on device performance. Conventionally, these species are detected and imaged using the electron spin resonance (ESR) technique. In many instances, ESR is not sensitive enough to deal with miniature devices having small numbers of paramagnetic species and high spatial heterogeneity. This limitation can in principle be overcome by employing a more sensitive method called electrically-detected magnetic resonance, which is based on measuring the effect of paramagnetic species on the electric current of the device while inducing electron spin-flip transitions. However, up until now, measurement of the current of the device could not reveal the spatial heterogeneity of its paramagnetic species. We provide here, for the first time, high resolution microimages of paramagnetic species in operating solar cells obtained through electrically-detected magnetic resonance. The method is based on unique microwave pulse sequences for excitation and detection of the electrical signal under a static magnetic field and powerful pulsed magnetic field gradients that spatially encode the electrical current of the sample. The approach developed here can be widely used in the nondestructive three-dimensional inspection and characterization of paramagnetic species in a variety of electronic devices.

## Significance statement

The detection and imaging of point defects, dopants, and impurities in solid-state devices is significant because of their effect on device performance. Conventionally, these species are observed using electron spin resonance, which suffers from limited sensitivity and spatial resolution. Recently, an alternative and much more sensitive detection method has emerged, based on measuring the effect of these so-called paramagnetic species on the device's electric current. However, this electrical detection method could not be used to obtain high resolution images of paramagnetic species in heterogeneous samples. We present here a recent methodological development that provides, for the first time, high-resolution microimages of the paramagnetic species in an operating device (solar cell) obtained through a combination of pulsed electrical detection and MRI-like imaging protocols.

## 1. Introduction

Paramagnetic species are an inseparable and essential part of any solid-state device, from the simplest diode through solar cells and up to the most complicated three-dimensional chip. Many paramagnetic species are intentionally inserted into the material (e.g., phosphorus and boron dopants in semiconductors) while others represent unwanted by-products, such as point defects and crystal impurities. Thus, both desirable and undesirable paramagnetic species constitute an inevitable part of the electronic devices' industry and greatly affect their performance (1). Traditionally, the identification and study of these species are carried out using electron spin resonance (ESR). For example, in the case of point defects and impurities in semiconductors, ESR enables researchers to characterize the defects' atomic structure, learn about impurity concentrations, and bridge the gap between paramagnetic and electronic properties. However, limited sensitivity and image resolution prevents ESR from being a major player in the nanoworld. ESR cannot be applied to small heterogeneous samples, such as common electronic devices that are often characterized by deep-submicron dimensions. Therefore, high sensitivity/high resolution ESR carried out in this length scale can open the door to a whole new world of applications. For example, ESR may be used for the direct study of point defects and defect clusters, their distribution, size, and to measure their diffusion or migration processes (see our initial work on this subject (2, 3)). ESR can also be used to correlate the device's electrical performances to defects, impurities, or dopant types and their exact physical location (for example, in the case of polycrystalline solar cells (4, 5)). This capability is especially important in the inspection of small devices, where average concentrations of such paramagnetic species measured in large bulk samples do not represent their actual local concentration in small dimensions - leading to significant variations in device performances (6). Thus, having ultra-high sensitivity/ultra-high resolution ESR would open a new era for electronic devices in which ESR could be used both as a purely basic scientific method and as a technological tool employed by device designers and for non-destructive three-dimensional failure analysis.

One approach recently adopted in order to resolve these problems involves the development of a ultra-high sensitivity ESR setup that makes use of miniature microwave resonators (7, 8). With our latest setup we have been able to achieve a sensibility of less than 1000 electron spins and spatial resolution of ~500 nm for a phosphorus-doped $^{28}$Si sample ($^{28}$Si:P). While this represents a significant advance over previous work, it still does not resolve all issues since the sensitivity (and the spatial resolution) strongly depends upon the relaxation properties of the species (known as spin lattice ($T_1$) and spin-spin ($T_2$) relaxation times). In that sense, the $^{28}$Si:P sample is quite optimal, but for others the spin sensitivity and spatial resolution are not as good.

Another possible approach to the ultra-sensitive detection of paramagnetic species in solid-state samples is based on measuring the electrical current in the device using the method known as electrically-detected magnetic resonance (EDMR)(9). This current was found to change when the electron spins of the paramagnetic species in the sample are subjected to a static field and microwave irradiation that comply with the resonance condition $\omega_0 \sim \gamma B_0$, where $\omega_0$ is the microwave frequency, $\gamma$ is the electron gyromagnetic ratio, and $B_0$ is the static field magnitude. The origin of this change in the electric current is traced down to the linkage of certain charge transport routes to the Pauli exclusion principle. There are several possible mechanisms that can result in a spin-dependent current. In the most general terms, resonant microwave irradiation will cause spin flips when set in the presence of an external magnetic field.. A charge carrier denied by the Pauli exclusion principle of the possibility to transit to another state and to sustain a current in the process may be permitted to undergo this transit once its spin has been flipped, or vice versa. The charge transport routes gated in this manner by magnetic resonance spin manipulations may be those associated with device inefficiency, such as leakage currents or premature charge annihilation in photo-voltaic devices (4), or they may exist as part of the desired device attributes (e.g., in organic electronics and spintronics devices (10)), depending on the type of the device and experimental conditions (applied voltage, temperature, etc.) (11-13). In this detection scheme, the energy quanta per spin is of the scale of eV, depending on the device's operating voltage, which is ~4-5 orders of magnitude larger than the energy quanta per spin in a typical ESR induction detection scheme. This leads to a much improved spin sensitivity of about 100 spins or better in a broad range of species (14, 15), thus dramatically surpassing the conventional spin sensitivity attainable via the conventional method known as induction detection. The EDMR method can be employed either with continuous microwave irradiation (CW-EDMR), or with pulsed microwave excitation (pulsed EDMR - pEDMR). The former is simpler and was developed already in the 1970s (9), while the latter is more complicated (developed only in 2001 (16)) and offers considerably more information about the relaxation properties of the paramagnetic species. Another important property of EDMR is that the changes in the electrical current provide direct information about the effect of paramagnetic species on the device's electrical performances.

A major factor that limits the applicability of EDMR, despite its abovementioned high sensitivity, is the fact that it cannot be applied to heterogeneous samples. This means that there is no useful capability that provides high-resolution spatially-resolved EDMR information. As already stated, this kind of information is vital to correctly analyze the factors affecting electronic device performances, such as microscopically inhomogeneous dopant concentrations or defect clusters, and to identify various types of defects originating at the boundary of microcrystals in microcrystalline-based devices, such as solar cells (4). In view of the importance of this type of information, much efforts have been recently invested in attempts to provide high-resolution spatially-resolved EDMR images using a conductive AFM tip (17). This method succeeded in providing a measurable signal even for a small electrode of ~3 μm in size. However, it suffers from significant current noise due to instability of the AFM contact; it requires the performance of sequential scanning of the surface; its current sensing mechanism is not necessary local; and it cannot be generalized to three-dimensional (3D) imaging requirements.

On the other hand, as noted above, conventional ESR using induction detection can provide high resolution 3D images of the sample of interest through the use of magnetic field gradients to spatially encode the sample signal. This image acquisition process is equivalent to the well-known medical imaging protocols used in MRI. ESR imaging can be employed either in CW or pulsed mode of gradients and data acquisition. In CW ESR measurements, spatial encoding is achieved by applying fixed magnetic field gradients during data acquisition (18) to spatially encode the sample. This approach is technically very simple but is very ineffective when the spectral linewidth of the sample is broad, leading to a very crude spatial separation. Thus, as an example, the only previously reported attempt to obtain EDMR-based images made use of CW EDMR data acquisition in a large silicon plate, under gradients of ~0.2 T/m, which resulted in an image resolution of ~1.9 mm (19). Clearly, such limited resolution is of no use to the vast majority of modern electronic devices of relevance and consequently this approach did not lead to any further insights. On the other hand, up until now it was not clear how pulsed ESR imaging protocols could be combined with an EDMR detection scheme, so this approach was abandoned.

Here we present a novel approach that makes use of a new pEDMR detection protocol combined with powerful pulsed magnetic field gradients to provide EDMR images of the paramagnetic species in operating thin-film solar cells with a spatial resolution of ~22 μm. The resolution in our experiment is limited only by the sensitivity of the detection. Thus, since the imaging protocol is based on our previous work with conventional pulsed ESR detection, the same setup can in principle achieve deep submicron and even nanoscale resolution, provided that the noise in the detection could be further reduced (which depends on the specific setup and type of sample measured). Furthermore, the new EDMR imaging scheme retains the ESR information regarding the paramagnetic spectrum of the measured species, thereby enabling their spatially-resolved assignment and characterization. Therefore, our line of work could lead to a new type of analysis tool for solid-state electronic devices with high spatial resolution.

## 2. The Pulsed EDMR Microimaging Setup

First, let us briefly describe the principles of pEDMR and then show how it can be generalized to include imaging capability. The most basic pEDMR detection scheme is shown in Fig. 1a (16). In pEDMR the observable is a transient current, $\Delta I$, induced following intense microwave (MW) pulses. In solar cells the pEDMR signals may have positive or negative signs, depending on the underlying transport process and operation conditions (see e.g. (13) for an overview on spin dependent transport processes in solar cells).

Another possible sequence used in pEDMR is the electrically-detected electron spin echo (ED-ESE), which is an extension of the standard two-pulse Hahn echo sequence with an additional $\pi/2$ readout pulse at the time of echo formation (see Fig. 1b) (20). The last MW pulse rotates the spin system back into singlet or triplet states, thereby transferring electron coherence to polarization. By integrating $\Delta I$, following the third MW pulse, over time one obtains the echo amplitude. To strobe the whole spin echo, the pulse sequence has to be iterated for varying $\tau_2$ times (nevertheless, for imaging, acquiring a single echo point is enough). In addition, as we discuss below, the imaging scheme we employ require having the full complex information about the spins' magnetization (21), i.e., both the I (in phase) and Q (out of phase) components of the magnetization as it precesses in the XY laboratory plane (with the Z-axis along the static magnetic field). In order to obtain these quantities we introduce here a new scheme where this ED-ESE sequence is repeated twice. Once with the detection pulse with X phase and then with Y phase. The first type of detection pulse flips to the Z-axis only the MW-affected spins that were positioned prior to the pulse along the Y-axis, while the second type of detection pulse flips to the Z-axis the magnetization that was positioned along the -X axis. The combination of these two measurements provides the full complex data about the spins precessing in the XY-plane, at the time of the echo.

Now, let us see how this detection scheme can be generalized to include imaging capability. Generally speaking, to obtain an ESR image, magnetic field gradients have to be applied to spatially encode the signal coming from the sample (22). As noted above, a fixed magnetic field gradient combined with CW EDMR detection is of no practical use, mainly due to the wide inhomogenously-broadened spectral line of most samples of relevance. On the other hand, previous research of EDMR with a variety of samples has clearly shown that it is a coherent phenomenon, exhibiting behavior that includes Rabi oscillations and echo refocusing (23). These characteristics can in principle be exploited to obtain high-resolution ESR images making use of pEDMR detection combined with so-called pulsed phase gradients for spatial encoding (21, 24). In this imaging method spatial information is encoded in the ESE phase angles by transient magnetic field gradients, as sketched in Figs. 1d and 1e in temporal and spatial domain, respectively. Gradients are applied during the first delay ($\tau_1$) of the ESE sequence. The gradient strength ($G_0$, $G_1$, ...) is incremented step-wise and for each step the ESE amplitude and phase is recorded. To illustrate the phase gradient imaging principle, their impact on electron spins located at three different points in space ($x_1$, $x_2$, $x_3$, respectively) is shown in Fig. 1f and 1g with respect to circles indicating rotating frames of reference (one-dimensional example). If no magnetic field gradient is applied ($G_0$) all three electrons spins will remain stationary in the rotating frame (Fig. 1f). In case of a finite gradient ($G_i$), electron spin $j$ acquires a phase angle $\varphi_{ij} = \gamma x_j \int\limits_{\tau_G} G_i dt$, which results in I and Q components of the ESE intensity of $I \sim \sum_j \cos(\varphi_{ij})$ and $Q \sim \sum_j \sin(\varphi_{ij})$ (Fig. 1g). Thus, in the general case, as a result of the phase gradients the ESE intensity and phase will oscillate as a function of $G_i$.  As depicted in Fig 1h, a Fourier transform (FT) of this oscillating signal with respect to $G_i$ yields the real space distribution of the electron spins.

The practical realization of such imaging scheme requires to overcome some major experimental issues of concern.  For example, pEDMR makes use of very sensitive current measurements of the sample under test.  However, the MW pulses and especially the magnetic field pulses create large transient current along the wire leads.  For canceling this disturbance we implemented a two stage solution that includes both MW phase cycling and the application of a "field jump" protocol.  Phase cycling is based on a ± phase modulation of the first MW pulse in the sequence (see Fig. 1b) (25), which in turn modulates the phase of the echo signal.  By repeating the sequence once with +X phase and then with -X phase, and then subtracting the results, the current transient is eliminated while the coherent EDMR echo signal that follows the phase of the first pulse, is reinforced.  The "field jump" protocol provides yet another mechanism for the reduction of the current transient (see Fig. 1c).  Here, again the current measurement is repeated twice, once "on resonance", and then "off resonance" by applying a fast current pulse through an auxiliary coil in the imaging probe that quickly (within a few μs) changes the resonance field.  The unwanted current artifacts are the same under both conditions and can thereby be eliminated by subtracting "off resonance" from "on resonance" current transients.  Thus, by applying this two stage solution (phase cycling and field jump) spin-dependent and spin-independent electrical responses may be separated from each other.

The novel pulsed EDMR imaging setup that was developed in this work is schematically described in Fig. S1.  The measurement setup is based on our conventional pulsed ESR microimaging system (26) together with our cryogenic pulsed ESR probehead (27).  In order to facilitate the generation and acquisition of the pulse sequence described in Fig. 1 we had to implement some software and hardware modifications in the system and in the cryogenic probehead, respectively.  Fig. 2 depicts a drawing of the cryogenic pEDMRI probe together with a zoom in photograph of the resonator and the mounted solar cell. The dielectric resonator is based on a double-stacked ring structure made of DR80 material from TCI Ceramics, Inc. The quality factor of the resonator is ~50, as measured with a vector network analyzer. For EDMR detection we extended the cryogenic probe (see Fig. 2), with an optical fiber, to illuminate the solar cell, by a halogen lamp or, alternatively, a green laser, and two shielded coaxial electrical leads, to supply the solar cell with a voltage bias and to measure the photocurrent. MWs are supplied via a coaxial MW feed line, that goes into the gradient coil fixture, where it is turned into a microstrip line, through an appropriate adapter. The microstrip line then goes below the resonator structure and excites the double stacked resonator by capacitive coupling. Its position can be varied with respect to the resonator by XY-stages to match the resonator impedance to that of the transmission line. In this position the solar cell is also connected to the coaxial wire leads and can be illuminated through the optical fiber.

With respect to the detection system (see Fig. S1), we added a sensitive current preamplifier for EDMR detection (28). In order to sample properly the in-phase and out-of-phase components of the magnetization, we made use of a computer-controlled diplexer switch (Mini-Circuits' model ZYSW-2-50DR), which directs the EDMR signal to the appropriate channel (I or Q) of the analog digitizer card in the computer. In addition, we implemented a software upgrade that properly controls EDMR spectroscopy and imaging sequences. This novel set-up now provides the unique possibility to perform coherent pEDMR spectroscopy and imaging on fully processed electronic devices in a single set-up, with unmatched spatial resolution.

### 3. Results and discussion

Before embarking on the actual EDMR microimaging experiments, we performed pEDMR measurements to assign spin-dependent charge transport processes in the solar cell, identify dominating noise sources that limit the detection sensitivity of the EDMRI spectrometer, and measure the spin coherence time.

Fig. 3 presents field-swept ED-ESE spectra recorded on an illuminated a-Si:H solar cell, following selective e-beam irradiation in its central part (see Materials & Methods section), at 100 K and 10 K, respectively. Spectra were recorded with the echo sequence depicted in Fig. 1b, without applying a field gradient or field jump. They differ with respect to their sign, the resonance position, and line shape. These differences originate in the fact that in a-Si:H at different temperatures different spin dependent transport processes dominate (29-31). The current enhancing signal measured at 10 K is a superposition of a narrow resonance at g = 2.004 and a broad line at g = 2.01. This pEDMR spectrum may be assigned to two independent spin-dependent hopping processes among conduction band tail states (cbt, g = 2.0044 (32)) and valence band tail states (vbt, g = 2.01 (29, 32)), respectively.

Upon increasing the temperature to 100 K the sign of the spin-dependent current is flipped and the main contribution to the spectrum is shifted to g = 2.005. The resulting EDMR spectra originates from spin-dependent tunneling of trapped cbt electrons into neutral Si dangling bonds (db, g = 2.0055 (29, 33)). At 100 K this signal still has some contribution from vbt states, which disappears above 200 K (see SI). The assignment of spin-dependent transport processes was further corroborated by X-band pEDMR measurements under varying ambient conditions (see SI). Results from ED-ESE spectroscopy will in the following set the basis for the interpretation of pEDMRI images recorded at 10 K and 100 K.

From the data depicted in Fig. 3 we obtained relative EDMR-induced current changes (relative to the DC current of the cell) $\Delta I/I \sim 9.25 \times 10^{-6}$ and $\Delta I/I \sim 0.0022$ at 100 K and 10 K, respectively. The single -shot noise RMS level was ~9 nA (100 K) and 4.5 nA (10

K).  This is roughly ~4 and 15 times higher than the predicted shot noise level (assuming a bandwidth of detection of 200 kHz.  The single-shot signal-to-noise-ratios were 0.087 (100 K) and 0.43 (10 K).  In the 100 K measurement this is probably due to noise contributions from the device itself and instabilities in the intensity of the halogen light source.  In the 10 K measurement the solar cell has much less noise and the light source instability may be identified as the dominant noise source.  In both cases, thermal noise is not significant.

Proper setting of the imaging pulse sequence requires prior knowledge about phase memory times, $T_m$.  By varying the pulse separation time $\tau_1$ from 0.5 µs to 4 µs we were able to estimate the EDMR detected $T_m \sim 3.6$ µs, which was found to be the same at 10 K and 100 K.  This $T_m$ is long enough to fit in the phase gradients between the 90° and 180° MW pulses, without excessive loss of echo signal.

Based on the parameters obtained from ED-ESE spectroscopy we proceeded to perform two-dimensional pEDMRI on as-deposited and e-beam degraded a-Si:H solar cells.  Fig. 4a shows a photograph of the a-Si:H solar cell alongside with the illumination profile.  In Fig 4b the pEDMR image of the as-deposited solar cell measured at 10 K under illumination is overplayed on the photograph of the solar cell.  For pEDMRI we chose the same MW pulse parameters as for ED-ESE spectroscopy.  In addition, we applied 400 ns long magnetic field gradients and the field jump protocol, outlined in Fig. 1c-d.  With these parameters and an accumulation time of ~4 hours pEDMR images with 100×64 pixels were obtained.  The pEDMR image depicted in Fig 4b is somewhat distorted, due to non-uniform gradient magnitude (which in principle can be corrected by off-line image analysis).  Nevertheless, the pixel resolution of ~22×34 µm clearly resolves the boundaries of the solar cell and the modulation of spin-dependent transport over the cell.  This modulation is due to the excitation profile of the light source, which doesn't illuminate the solar cell uniformly.  The spatial distribution of the spin-dependent transport signal over the cell may be rationalized by the underlying process.  At 10 K and with applied bias the pEDMR signal is dominated by spin-dependent hopping via cbt and vbt states, which increases with increasing light intensity incident on the solar cell.

In the following, we carried out pEDMRI on a solar cell, which was subjected to e-beam degradation within a 500 μm × 500 μm diamond shaped region (see Materials & Methods). Figs. 4c and 4d depict pEDMR images obtained from the irradiated cell at 10 K and 100 K, respectively. Except for the number of pixels (64×40 for 100 K and 80×60 at 10 K) these images were acquired with the same experimental parameters as Fig 4b.

Comparing 10 K pEDMR images obtained on as-deposited (Fig. 4b) and e-beam degraded (Fig. 4c) solar cells, following spin-dependent current patterns may be identified. In both cases the strength of the spin-dependent hopping signal depends on the excitation profile of the light source. However, in addition the e-beam degraded cell shows a pronounced decrease of the spin dependent current as compared to the as-deposited cell. This interesting finding is not fully understood yet. Possible reasons could be an increase of non-spin dependent recombination pathways via doubly occupied dangling bonds (29) or a reduction of paramagnetic tails states through shifts in the a-Si:H Fermi level (31).

Upon increasing the temperature the pEDMR image of the solar cell again changes dramatically (see Fig. 4d). At 100 K the e-beam irradiated region exhibits a strongly increased EDMR signal as compared to the surrounding parts of the solar cell. From EDMR spectroscopy we found that the 100 K EDMR signal originates from recombination involving cbt states and dbs. Increasing the number of dangling bonds by e-beam degradation, therefore, leads to increased spin-dependent recombination in this region.

These first high resolution functional pEDMR images on operating solar cells demonstrate the potential of this novel technique to locate spin-dependent transport and loss mechanisms in an electronic device and conclude on the distribution of function determining paramagnetic states in fully processed electronic devices.

## 4. Conclusions and future prospects

The basic approach to pulsed EDMR imaging by pulsed phase gradients has proven to work well. Nevertheless, there is still plenty of room for improvement. The fundamental limiting factor in our EDMR imaging experiment is the signal-to-noise-ratio (SNR). Our phase gradient drivers can provide much more powerful gradients than the ones used in this experiment and currently support resolutions down to 80 nm (26, 34). However, at such resolution level the noise in our experiment will be too high, since the number of defects, or states, in a given voxel would be too small to observe. Improvement in SNR can be achieved by increasing the light's intensity to at least 3.2 suns (leading to ~16 times more current than obtained here, meaning an increase in SNR by a factor of 4), and using a more stable light source that does not have significant noise component at the ~1-500 kHz range (this can increase SNR by a factor of ~4 by reaching the shot noise limit). Additional significant improvements can be gained by using smaller cells with a smaller overall shot noise (e.g., a cell with a size of ~100×100 $\mu$m would lead to a tenfold improvement in SNR).

The cumulative effect of these near-future improvements can increase SNR by a factor of ~160, meaning that it is possible to reach an image resolution down to almost 1 $\mu$m for the type of cells we employed here. We can estimate the concentration of the conduction band tail states that contribute to the spin-dependent current component in this material to be ~$10^{16}$ states per $cm^3$ (35). This means that Fig. 4 shows ~$10^7$ states in each voxel in the EDMR image, with an SNR of ~200. Other solar cells or semiconductor devices that have paramagnetic species or states with larger concentrations may make it possible to obtain even higher spatial resolution in the nanometer scale. For example, since an EDMR signal for P-doped Si can be obtained for less than 100 spins, it means that a 3D resolution of ~100 nm should be readily available with such type of sample having a P concentration of ~$10^{16}$ atoms/$cm^3$.

The method developed here can be of wide use for the nondestructive inspection of paramagnetic species in a variety of semiconductor devices. Although the current results have limited resolution, relatively simple future improvements can greatly enhance the capabilities of the pulsed EDMR experiment. It should also be noted that our system can in principle support also 3D and 4D imaging capabilities (with the 4[th] dimension referring to the spatially-resolved EDMR spectrum). This can be of importance to the emerging field of 3D semiconductor devices.

## 5. Materials and Methods

EDMR measurements were carried out on thin-film solar cells (see Fig. S2) with 1000 nm thick hydrogenated amorphous silicon (a-Si:H) absorber layers sandwiched between microcrystalline silicon ($\mu$c-Si), p and n layers and transparent top and bottom contacts made from Al doped ZnO. Solar cell samples were deposited on quartz substrates by plasma enhanced chemical vapor deposition in superstrate configuration (36), which allows illumination through the substrate.

In order to modulate the spatial distribution of dangling bond defects by high energy electrons (37, 38), some of the solar cells were exposed to the 20-keV electron beam (e-beam) of a scanning electron microscope through the top contact opposite to the substrate (top side in Fig. S2b). An electron current of ~13.5 nA and a dose of 10 mCb/cm$^2$ was applied at a beam energy of 20KeV to engrave a diamond shaped pattern (500×500 $\mu$m, within dashed diamonds in Figs 4c and 4d into the cell). Throughout the manuscript we refer to e-beam treated samples as e-beam degraded solar cells, whereas the untreated are labeled as as-desposited solar cells.

## 6. Acknowledgements

This work was partially supported by grant # G-1032-18.14/2009 from the German-Israeli Foundation (GIF), grant #213/09 from the Israeli Science Foundation, grants #201665 and #309649 from the European Research Council (ERC), and by the Russell Berrie

Nanotechnology Institute at the Technion. MF received funding from the German Research Foundation within SPP 1601.

## 8.  Figure captions

Figure 1:   (a) Basic field swept simple one-pulse pEDMR sequence, (b) ED-ESE MW sequence with the ED-ESE plotted vs. $\tau_2$.  $\pm$X and X/Y indicate the MW phases of the first and third pulse, respectively,  (c) timing of the magnetic field jump, that switches between on and off EDMR conditions, (d) timing of the transient magnetic field gradients relative to the MW pulses, (e) spatial magnetic field offset with ($G_1$) and without ($G_0$) field gradient, (f) and (g) circles indicating rotating frames of reference (one-dimensional example) for electron spins (red arrows) located at three different points in space ($x_1$, $x_2$ and $x_3$) without and with applied field gradient, respectively, (h) oscillating ESE intensity vs. field gradient.

Figure 2:   (a) Drawing of the cryogenic pEDMRI probe with electrical and optical supply lines, the gradient coil array and the resonator insertion module with mounted solar cell.  (b)

A zoom-in photograph indicating the position of the solar cell inside the double-stacked resonator (ring dimensions: o.d. = 4.4 mm, i.d. = 1.5 mm, height = 2.3 mm, distance from pair ring = 1.4 mm). Overlaid on the photograph, directions of the external static magnetic field ($B_0$), the magnetic field component of the microwave ($B_1$) and the illumination (when inserted into the gradient coils) are shown.

Figure 3: ED-ESE spectra of an illuminated a-Si:H solar cell at 10 K (top) and 100 K (bottom), respectively. Absolute ED-ESE intensities were obtained with the three pulse mw sequence described in Fig. 1b. Alongside with the experimental spectra (blue solid lines), spectral simulations (red dashed lines) assuming contributions from conduction band tail (cbt) states (green dashed lines), conduction band tail (vbt) states (grey dashed lines), and dangling bond states (dbs) (pink dashed lines) are plotted, with their respective g values given in the legend. Experimental parameters: MW frequency = 8.3 GHz, light intensity = ~0.2 suns (200 W/m$^2$), $\tau_1$ = 400 ns, MW pulse lengths were $\pi/2$ = 70 and $\pi$ = 130 ns, bias voltage over the cell -0.74V, DC current under illumination $80 \times 10^{-6}$ A and $0.9 \times 10^{-6}$ A at 100 K and 10 K, respectively, accumulations per field point 50,000, and repetition rate 20 kHz.

Figure 4: (a) Photo of the solar cell and its components, showing also the illumination area. (b) pEDMR image of the illuminated as-deposited solar cell at 10 K, when measured inside the double-stacked dielectric resonator. (c) and (d) pEDMR images of the e-beam degraded solar cell, measured at 10 and 100 K, respectively. The degraded area is framed by the dashed diamond shape. The color code appearing in (d) is linearly scaled and normalized to the largest pixel signal in each image.

Experimental parameters: MW frequency = 8.3 GHz, light intensity = ~0.2 suns (200 W/m$^2$), $\tau_1$ = 500 ns, MW pulse lengths were $\pi/2$ = 60 and $\pi$ = 120 ns, magnetic field gradients $\tau_G$ = 400 ns, maximum strength 6.5 T/m (for the X axis) and 4.3 T/m (for the Y axis), bias voltage over the cell -0.74V, DC current under illumination $70\times10^{-6}$ A and $0.8\times10^{-6}$ A at 100 K and 10 K, respectively, and repetition rate 20 kHz and 4kHz respectively.

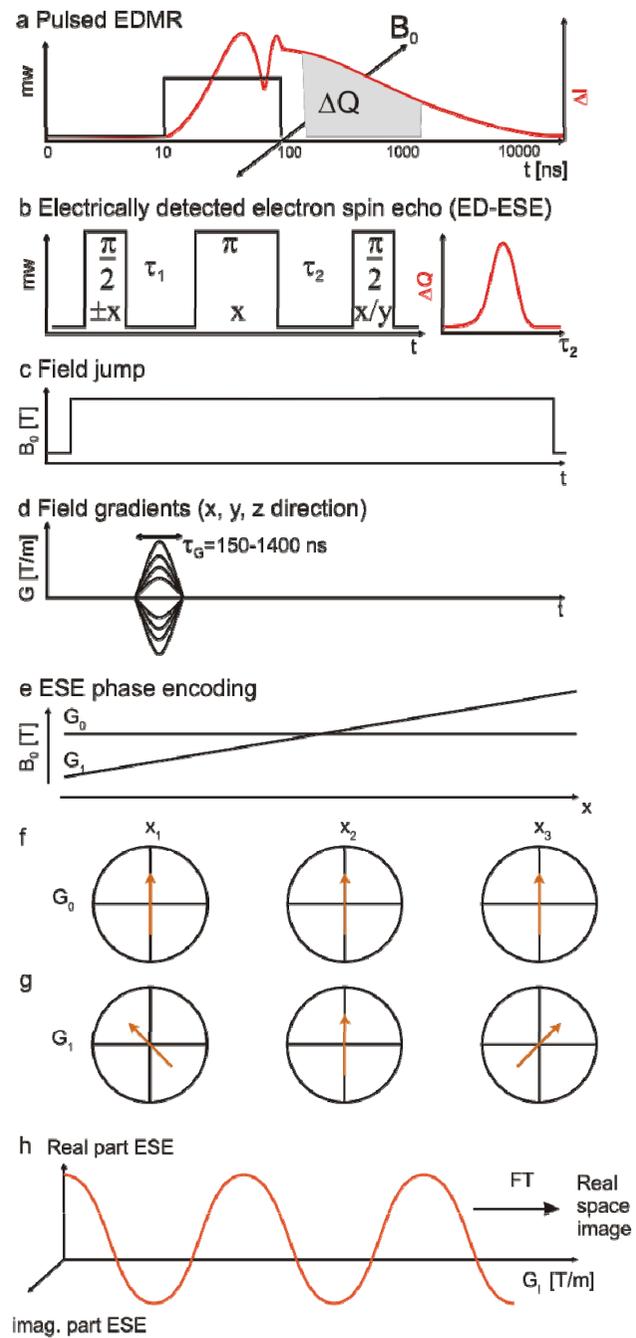

Figure 1

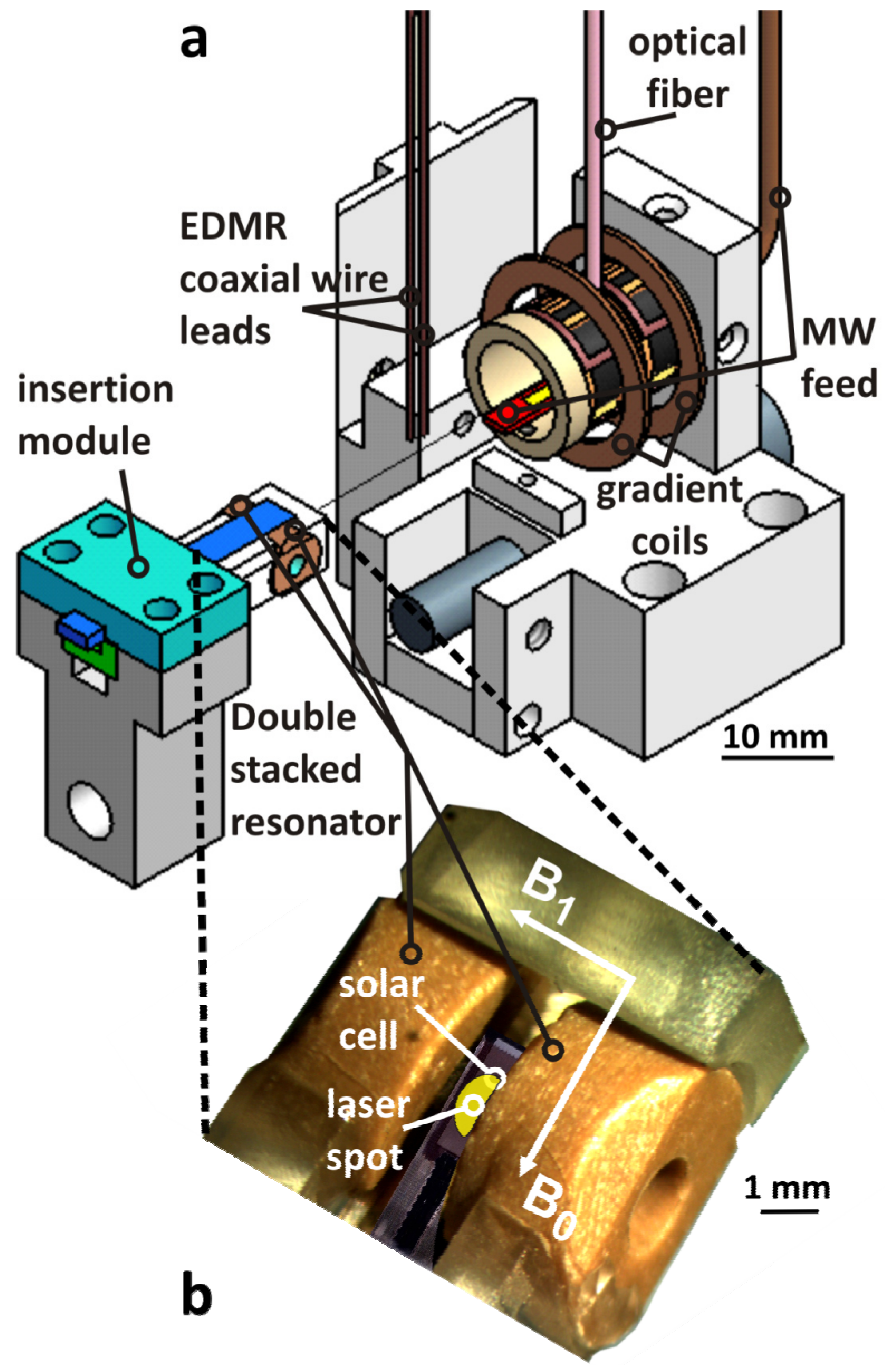

**a**

optical fiber

EDMR coaxial wire leads

MW feed

insertion module

gradient coils

10 mm

Double stacked resonator

B₁

solar cell

laser spot

B₀

1 mm

Figure 2

**b**

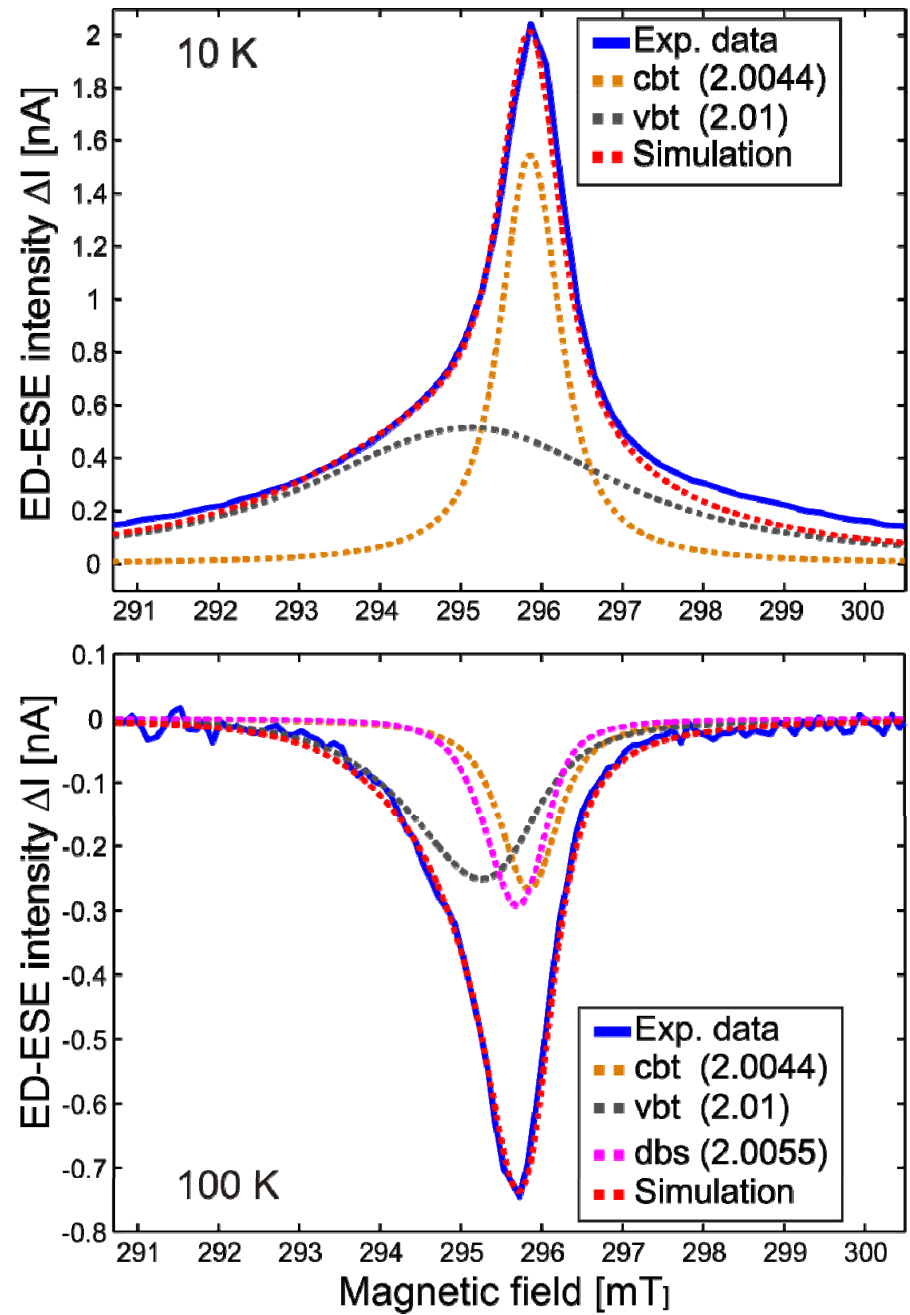

Figure 3

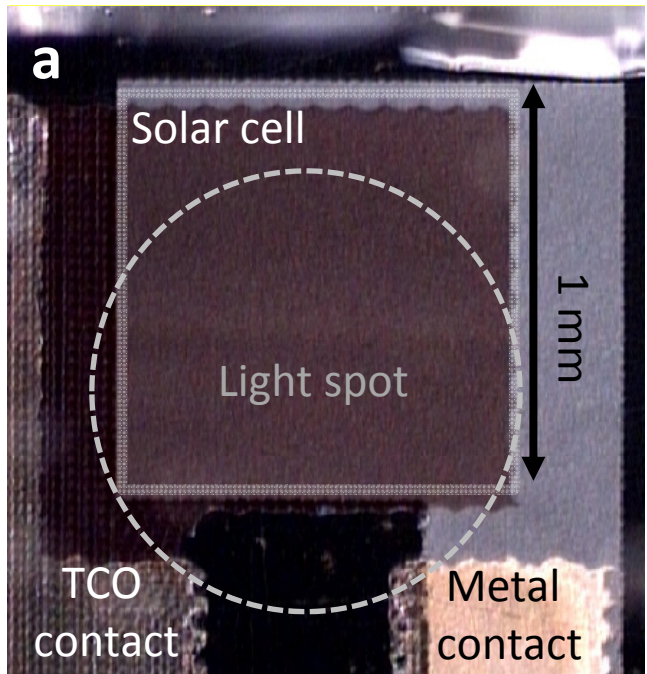

**a-Si:H solar cell**

a

Solar cell

Light spot

1 mm

TCO contact

Metal contact

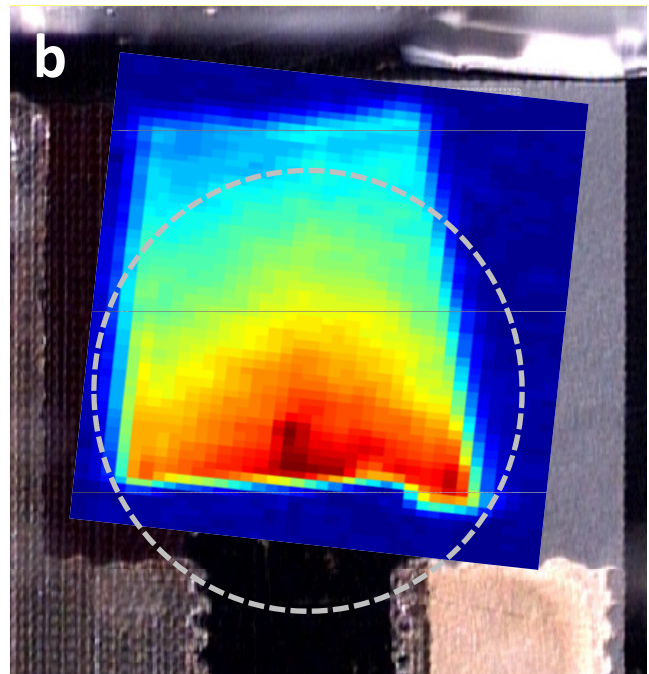

**EDMRI, T=10 K**

b

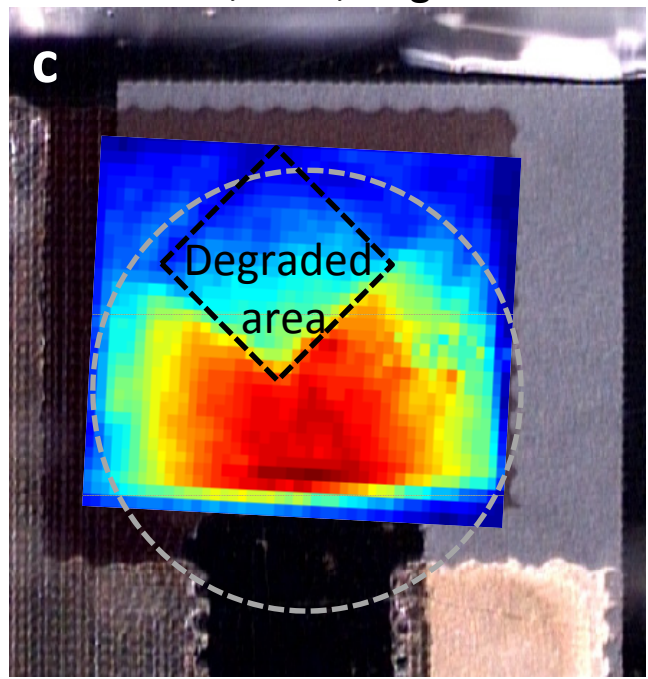

**EDMRI, 10 K, degraded**

c

Degraded area

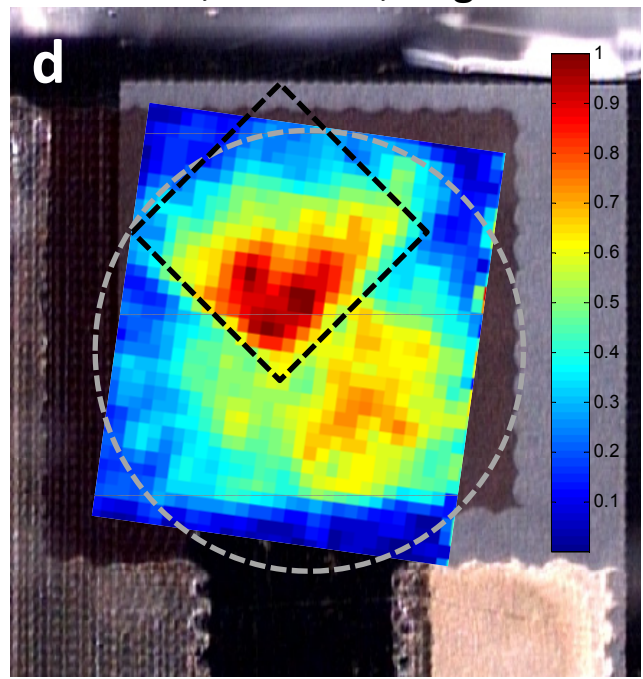

**EDMRI, T=100 K, degraded**

d

Figure 4

# High Resolution Microimaging with Pulsed Electrically-Detected Magnetic Resonance

## Supplementary Information

### 1.    The EDMR imaging system

Figure S1 shows the block diagram of the EDMR microimaging system.    The measurement setup is based on our conventional pulsed ESR microimaging system (full details can be found in refs (S1) and (S2)), to which  certain hardware and software features have been added, as noted in the main text.

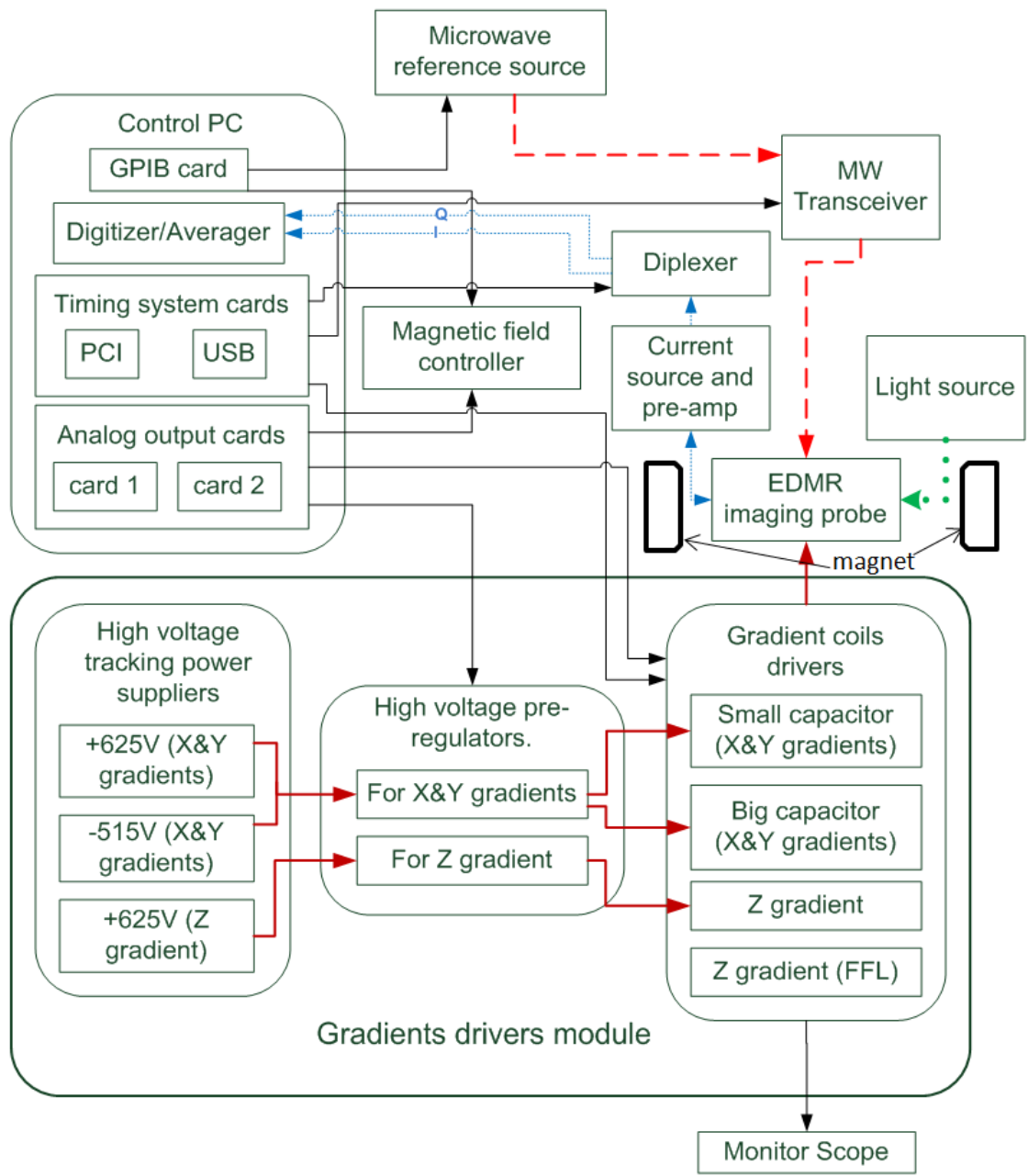

## 2.    The solar cell samples

Figure S2 describes the structure of the thin-film silicon solar cells that were used in this work.

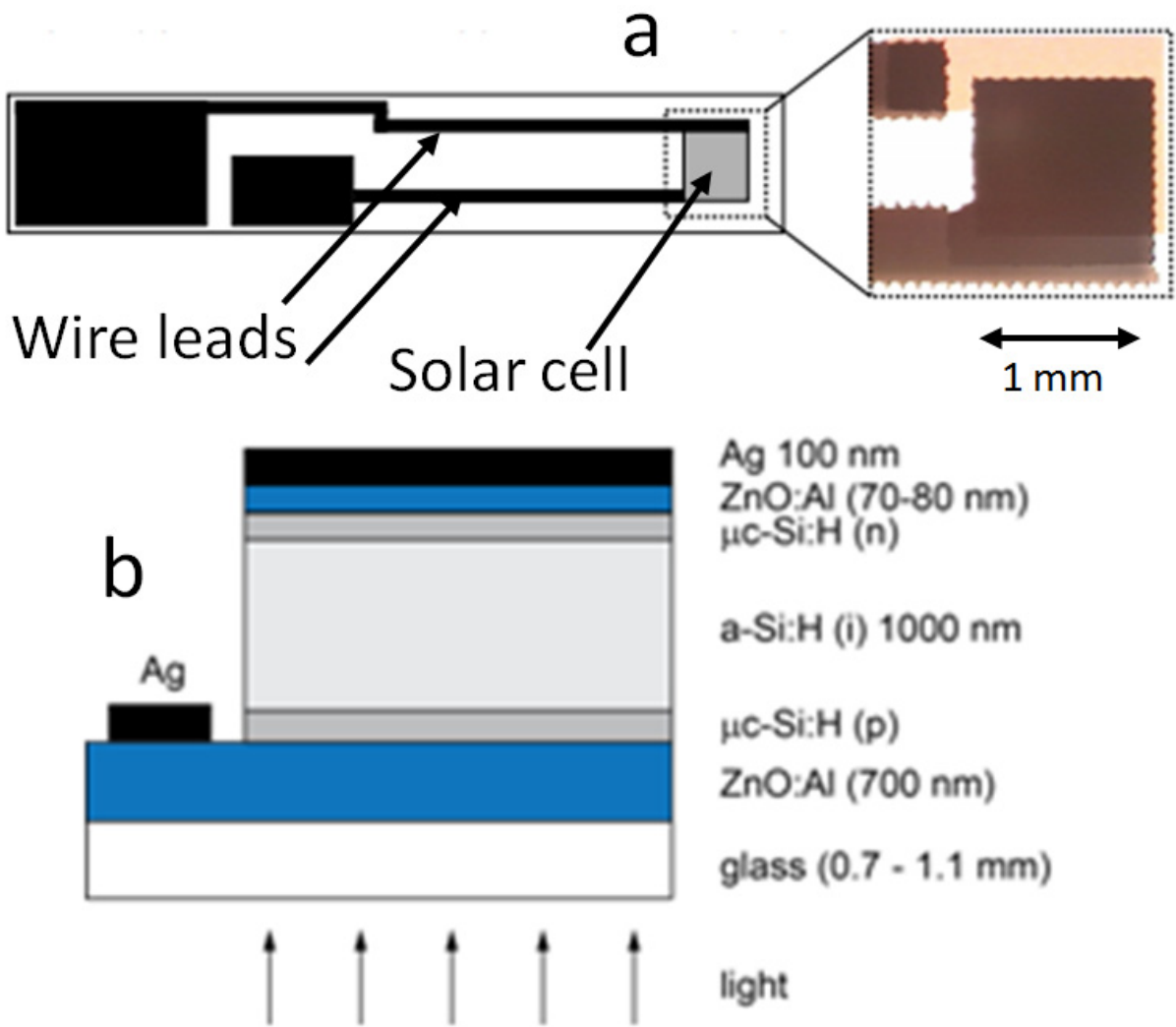

Figure S2: (a) General layout of the a-Si:H thin-film silicon solar cell used in this study.  A close-up of the solar cell is shown on the right side.  (b) Lateral structure of the cell, light is entering through the glass substrate.

## 3. Identifying the sources of the EDMR signal

The identification of the sources of the EDMR signal in our solar cell samples was carried out by combined CW and pulsed spectroscopic EDMR measurements, performed on a standard X-Band Bruker Elexsys spectrometer at the Helmholtz Center in Berlin.

The CW-EDMR measurements of the sample after e-beam degradation are shown in Fig. S3. The measurements were performed under short-circuit conditions (U = 0 V) and under light illumination. All spectra were subject to post-processing to adjust the phase so as to minimize the quadrature (out-of phase) signal. The phase rotation angle which was applied to the individual spectra is plotted in S1b. A strong CW-EDMR signal could be observed at room temperature with a value of g = 2.0056. At lower temperatures, a signal with a value of g = 2.0046 was observed, which has a phase value larger by 203° than the CW-EDMR spectrum observed at room temperature. The phase shift and the observed g value indicate that the signal at room temperature can be attributed to a spin-dependent recombination through DB defects while the low-temperature signal is due to spin-dependent hopping among conduction band tail states.

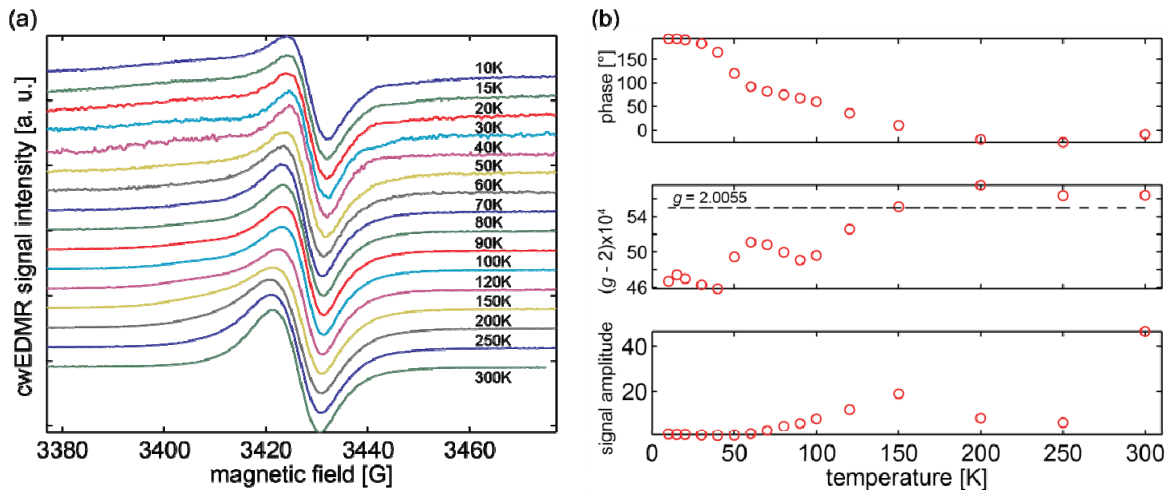

Figure S3: (a) Normalized CW-EDMR spectra as a function of temperature, showing real part. The magnetic field has been rescaled to the same MW frequency. (b) CW-EDMR phase, g value, and signal amplitude corresponding to data shown in (a). Measurement conditions: U = 0 V, illumination with halogen lamp (152 W), X-Band Bruker Elexsys (ER-4118X MD5 resonator), MW power = 10 dB, modulation amplitude = 4 G, frequency = 10 kHz.

Pulsed EDMR measurement were performed to investigate which spin-dependent transport processes dominate at different temperatures. The transient EDMR signal after a 120-ns microwave π pulse was measured. The results are shown in Fig. S4, where Fig. S4a displays the transient signal at the maximum of the spectrum and Fig. S4b shows the spectrum at a time t after the pulse, where the transient signal has reached a maximum (for a current enhancing signal) or minimum (for a current quenching signal). One advantage of the pulsed EDMR experiment over the CW-EDMR experiment is the possibility to directly determine whether the current decreases or increases on resonance. Figure S4 clearly shows that the signal at low temperatures (10 – 50 K) is a current-enhancing signal (the polarity of the current detection setup was chosen so that the total photocurrent flowing through the solar cell is positive). This indicates that spin-dependent hopping among conduction band tail states is observed at these temperatures. This conclusion is further supported by the measured value for g = 2.004, which is identical to the g value of conduction band tail states in a-Si:H (S3, S4). At higher temperatures (90 – 292 K) a strong negative transient EDMR signal is observed, which indicates that under these conditions a spin-dependent recombination through DB defects is observed. This is further corroborated by the value g = 2.0055, which is identical to the g value of dangling bonds defects (S4, S5). At intermediate temperatures it is difficult to distinguish clearly between hopping and recombination, because the transient EDMR signal becomes small and has both positive and negative contributions.

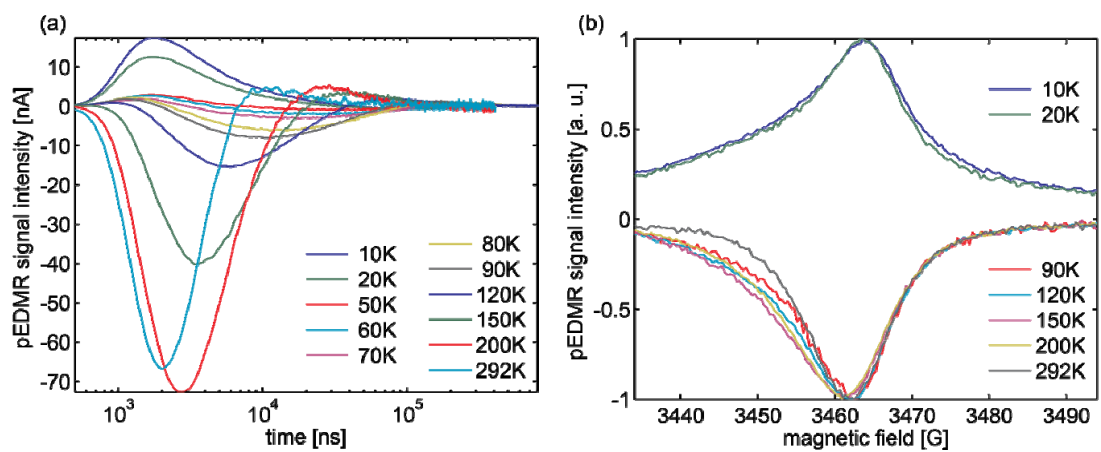

<u>Figure S4:</u> Pulse EDMR transients (a) and normalized spectra (b) as a function of temperature. Measurement conditions: U = 0 V, illumination with Halogen lamp (152 W), X-Band Bruker Elexsys (ER-4118X MD5 resonator), MW power = 18 dB, pulse length = 120 ns.

Additional CW EDMR spectroscopic studies, shown in Figure S5 and carried out at X-band and 263 GHz reveal a bit more complicated picture: While at low temperature there is indeed no evidence for dangling bond in the EDMR spectra, at higher temperature, the spectra is a mixture of both dangling bond and tail states (conduction and valance band). However, as noted above, in our pulsed EDMR data the dangling bond-related signal is the dominant contribution in the EDMR echo at higher temperatures (~ 100 K), while the tail states are the ones that play the main role at low temperatures (~10 K).

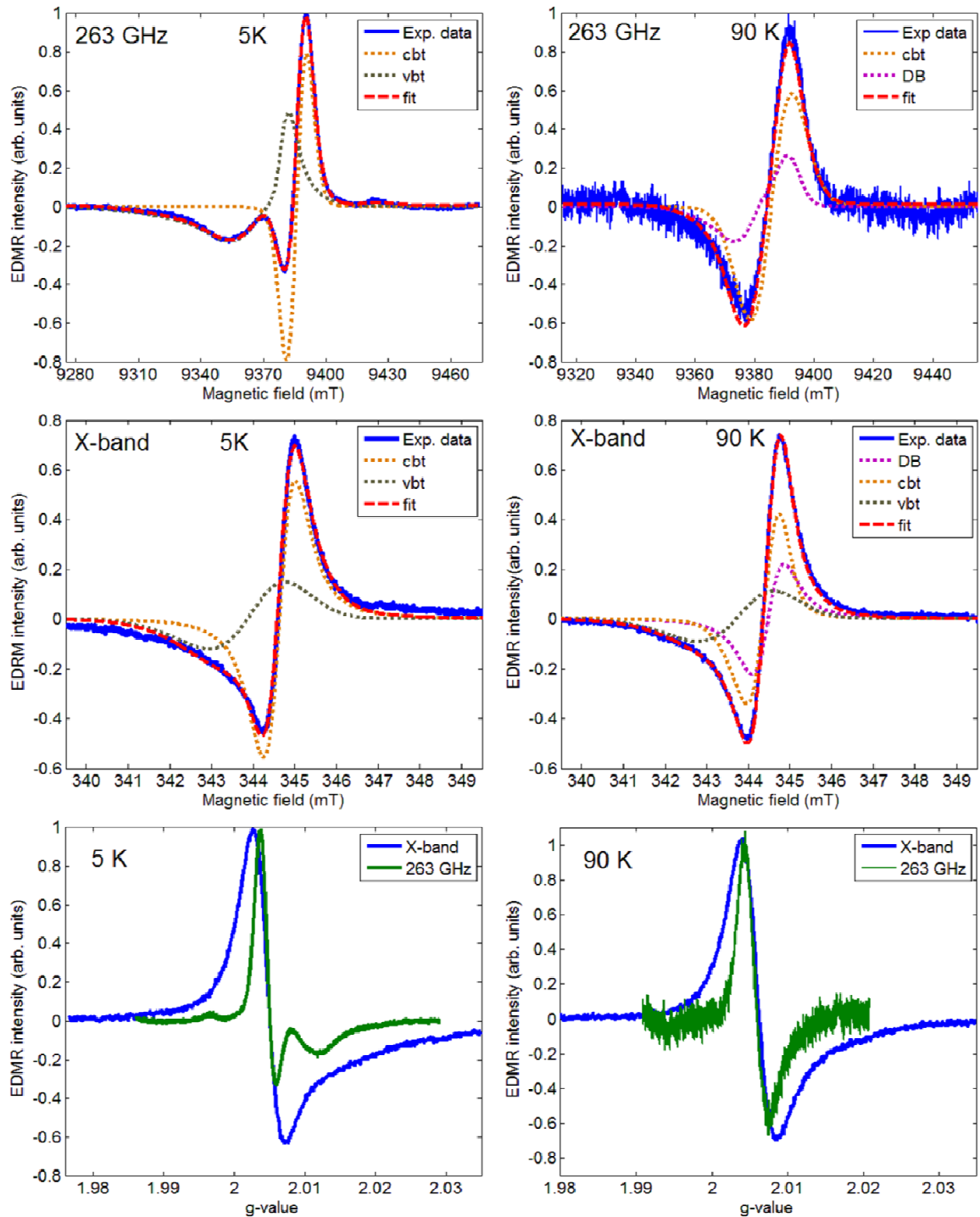

Figure S5: Spectroscopic CW EDMR studies of the solar cells at X-band and 263 GHz. The results show that the spectrum at 5 K is simulated with two main components, conduction band tail states (cbt) and valance band tail states (vbt). These states contribute also to the EDMR signal at 90 K, with the dangling bonds (DB) becoming also an important contributor to the signal.